\documentclass{aa}
\usepackage{graphicx}

\begin{document}
\title{Observational properties of visual binaries as modelled using a
synthetic catalogue}
\subtitle{I: For visual apparent magnitudes $m_v < 15$}
\author{Pasi Nurmi$^1$,\thanks{Pasi.Nurmi@oma.be} \and Henri M.J. Boffin$1,2$,\thanks{Henri.Boffin@oma.be}}
\institute{$1$ Royal Observatory of Belgium, Av. Circulaire 3, 1180 Brussels, Belgium\\
$2$ European Southern Observatory, Karl-Schwarzschild-str. 2, 85748 Garching-bei-M\"unchen, Germany}
\date{Received 31 January 2003 / Accepted 4 June 2003}
\abstract{
A simulated binary star catalogue is created that represents the full sky
distribution of stars, excluding those with galactic latitude $|b|<10\degr$,
down to $m_v < 15$. The catalogue is compatible with observed magnitude and
number distributions of the HST Guide Star Catalog. The simulated catalogue is
analyzed for observational properties of visual binary stars, both in a
general context without taking any observational constraints into account,
and using different observability criteria that model the observational
capability of the astrometric DIVA satellite. It is shown that several
observational bias effects should be taken into account when final output
catalogues are analyzed. The results indicate that $\Delta m_v$ distribution
peaks at $\Delta m_v \approx 7$ and to maximize the number of binary detections
with large mass ratios, the magnitude threshold should be set as high as
possible. This is especially important if low mass $M<0.2 M_{\sun}$
secondaries are studied. The predicted number of optical pairs is calculated
and for large magnitude differences close to the galactic plane and toward
the galactic center it may be difficult to discriminate between a red
companion and a low-mass foreground/background star for distant double stars
for which no parallax measurements are available. Using optimal detection
procedures for optical double stars it is estimated that, depending on the
assumed double star distributions, 60\%-70\% of all the visual binary stars in the 
catalogue with 
separations $0.0625\arcsec<\theta<3.5\arcsec$, $m_{v}(\mathrm{secondary})<22$, and 
$\Delta m<10$ 
could be observed during the DIVA or similar missions. 
\keywords{methods: data analysis
-- stars: binaries: general -- stars: binaries: visual}
}

\authorrunning{P. Nurmi and H.M.J. Boffin}
\titlerunning{Simulated catalogue of visual binaries}
\maketitle

\section{Introduction}
To optimize the forthcoming astrometric missions in their capabilities to
detect different types of binary stars 
requires detailed preliminary planning. One way to
do this is to create an artificial catalogue that has statistically similar
properties to the expected final output catalogue.
This allows us to do 
pre-launch fine-tuning of the methods and procedures used in the final data
reduction of forthcoming astrometric satellites. 
Also, some other binary studies using already existing catalogues
may benefit from the results of this kind of analysis.
With this aim in mind, we create an artificial binary\footnote{Throughout the
paper we use the word `binary' to describe a physical system while by double
star we refer to a system that can be a physical (i.e. gravitationally bound) 
or an optical
double.} catalogue that represents the overall distributions of  binary
stars in the sky down to visual magnitude $m_v=15$. This limit is chosen so
that it represents the observational limit of the DIVA satellite and it is also,
roughly, the completeness limit of the source
catalogue used, GSC 1.2 (the enhanced version of the Guide
Star Catalog 1.2; see \cite{morrison_01}) from which initial apparent magnitude
 distributions are taken. 
We should note that after the submission of this article 
all further development of the DIVA mission was stopped and DIVA
is now definitely canceled.
Our results will, however, still be useful for future similar missions but, more 
importantly, they represent a first necessary exploratory step in addressing the
possibilities of the GAIA astrometric mission with regards to the detection of 
binaries.

Our approach differs from traditional star count models (e.g.
\cite{bahcall_80},
\cite{gilmore_83}, 
\cite{bahcall_84}, etc.) by the fact that we use an
existing source catalogue directly as our observational basis and the galaxy model is
considered only when proper weights in the final distribution of binary
properties  are taken into account.   
In this study we consider only relative numbers with respect to the total
number of binary stars assumed to exist for all the stars in the sky down to 
$m_v \approx 15$. Absolute numbers could be calculated if the numbers given
in the paper are multiplied by the overall binary fraction $f_{binary}$ and
the total number of stars $N_{tot}$. 
According to Duquennoy and Mayor (1991) $f_{binary}$
is $\sim 0.6$ for G-dwarf field stars, 
but it may be different for cluster stars and different spectral types 
(Duch\^{e}ne et al. 2001). 
Using star counts from the GSC 1.2 catalogue 
and considering only those objects
having non-star classification, the absolute
number of stars can be estimated to be $\sim 2.1\times10^7$ stars in the
whole sky. 
This can be compared with the galaxy model prediction by Bahcall and Soneira
(1980)
that gives $3.7\times10^7$ stars when $m_v<15$.
The difference is mainly due to the incompleteness of GSC and the fact that 
some non-star classification objects are in fact close-by stars.

\section{The input for the binary catalogue}
Our approach is to create an artificial binary catalogue that is compatible
with 
\begin{itemize}
\item stellar evolutionary theory
\item initial mass function (IMF)
\item observed stellar densities of the sky
\item the assumed stellar populations in the galaxy
\item observed orbital distributions of binary stars
\end{itemize}
To fulfill these requirements we have used a hybrid approach in the 
binary catalogue construction. The main idea is to start from the IMF and
to evolve stars according to their age and metallicity distributions and
place them at random distances from the Earth. 
Then by varying the distance and stellar properties we find a match to the
observed $m_v$ distribution of GSC catalogue and repeat the whole procedure 
for every individual field and then collect the information covering the
whole sky. In the simulations typically $\sim 10^3$ random fields are chosen
each having a size of $\sim 1$ square degree. Thus, in the final
sample we have $\sim 4 \times 10^5$ binary stars, sufficient for our
statistical analysis. However, it should be noted that since we use random
distances and weights
in the final analysis, the number of hypothetical stars that would be
required to produce similar results using the galaxy model directly 
is much larger by several orders of magnitude.

\subsection{Initial stellar distributions and GSC 1.2 source catalogue}
In the simulation we use the GSC 1.2 catalogue as an input catalogue to have
a correct magnitude distribution and stellar density in each sample of the
sky. The nominal magnitude limit of the catalogue is 15.5, but it is not
fully complete down to this limit. 
To evaluate the completeness of the GSC we have
estimated the completeness by calculating the magnitude $m_{limit}$ 
at which $m_v < m_{limit}$ for 90\% of the stars in $10^3$ random fields: 
$m_{limit}$  corresponds roughly
to the turnover point of the cumulative distribution of stars.
The results are averaged over the galactic longitude $l$ and they are shown as a
function of ecliptic latitude $b$ (figure 1). 
It is evident that the completeness limit
is much smaller at low galactic latitudes. Since it is 
only close to the galactic plane that
the completeness is not so good,
we have not included
any corrections for this, since this has no effect on our
statistical results. This was also confirmed by doing a smaller scale simulation
using a collection of GSC 2.2 fields where the catalogue is complete 
to much larger magnitudes than that of the GSC 1.2. 
\begin{figure}
\centering
\resizebox{\hsize}{!}{\includegraphics[angle=-90,width=10 cm]{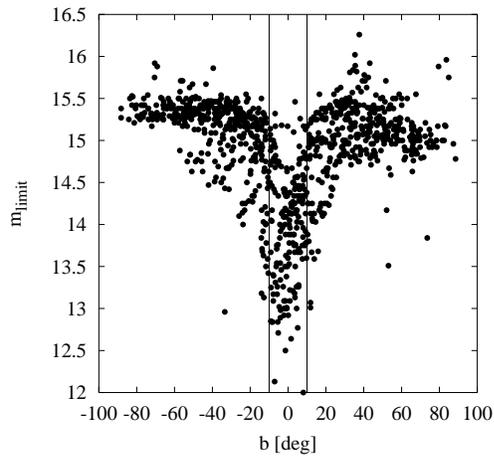}}
\caption{The magnitude limit of the cumulative distribution where 90\% of
objects have $m_v < m_{limit}$ as a function of $b$ averaged over $l$.  
The zone of avoidance in the calculation is
located between the two vertical lines.}
\label{}
\end{figure}

For our purpose the GSC 1.2 represents the full-sky distribution of stars 
well enough so that it gives suitable
observational data for the magnitude distribution during the magnitude fitting.
In GSC 1 mainly two band-passes are used in the plates:
photographic V (north) and photographic J (south) and short exposure V plates 
are used in the galactic plane.
Using the colour equation from Tritton (1983),
$\mathrm{J}=\mathrm{V}+0.72(\mathrm{B-V})$, and assuming 
that the difference between photographic V and Johnson V are small,
we have 
changed V and J magnitudes to
Johnson V bandpass. 
This approximation is good enough for our purpose, because
we are only concerned with a magnitude precision of 0.5 mag.
Since the colour is known from
the simulations, all the magnitudes are scaled  to $m_v$ using 
colour conversion equations given in Caldwell et al. (1993), 
where coefficients
of polynomial fits for different colour conversions are given.
Since we transform GSC magnitudes to $m_v$ and use this magnitude in our 
synthetic catalogue we need not to convert from $m_v$ back to GSC 
photometric magnitudes.

The overall magnitude errors of the catalogue are a few tens of
magnitudes. For our purpose this accuracy is high enough, since we are
only interested in statistical averages of the simulated catalogue.
The only requirement in the fitting procedure is that the magnitude
histogram is accurate enough at the one magnitude level.

The digital resolution of the scanning is 1.7\arcsec,
and all the binaries in our simulations 
that have separations $\theta$
less than this are considered unresolved and their combined magnitude is
used in the fitting procedure. For resolved binaries we use the magnitude of
the brighter star.

To calculate the observed $m_v$-distribution,
we select a large number of fields in the sky that are
transformed into galactic coordinates $l_c$ and $b_c$, where $c$ denotes the
central position of the field. A selected field is a circle with radius of
30\arcmin. The sample size should be large enough so that the surface stellar
density variations over the sky are well sampled, but making the field size
larger does not bring any additional information. 
The observed $m_v$-distribution is then calculated from this list of stars
after the conversion, explained above, is applied to the GSC magnitudes.

\subsection{Stellar evolution}
	
	    \subsubsection{IMF}
	    There is strong evidence that initial masses of stars in
different birthplaces all derive from the same initial mass distribution
(\cite{kroupa_02}). In our simulations, individual masses of stars 
are chosen from the initial mass function that is taken from Kroupa (2002).
By using the observed star count data and stellar luminosity functions Kroupa et al. (1993)
derived the IMF that is corrected for unresolved binaries. 
They assumed the binary fraction to lie between 0.6 and 0.7 and that binaries 
are derived from the same IMF as a result of random pairing.
The derived mass function for single stars is a multi-part power-law having turn-off points at
$0.5M_{\sun}$ and at $1M_{\sun}$. Brown dwarfs are not included in the simulation.

The detailed shape of the distribution
and the coefficients of the power-law are given in the supplementary material of the
Kroupa (2002) article.

Since the agreement about the IMF is not complete we have
also used the IMF of Miller and Scalo (1979) 
and the observed present day mass
function (PDMF) obtained for low-mass stars by Kroupa at al. (1990), 
but it turns out that the results remain roughly the same.
This is mainly due to the fact that functions differ mainly in their low and far
ends that are not statistically important in the final observed 
distributions.
The masses used in the simulations are between $0.08M_{\sun}$ and 
$100M_{\sun}$. These are also the limits of applicability of
Hurley et al. (2000) stellar evolution algorithms.

However, it is not clear how binaries should be formed from the IMF
or if a general IMF for binaries does exist at all. In
principle there are three different possibilities to create a binary from the
initial IMF. The
most simple way, as used in the IMF derivation obtained by Kroupa et al. (1993), 
is just to choose two random masses from the same IMF and 
create a pair. 
Another way is to choose a combined mass from the IMF and use
some distribution for the mass ratio $Q=m_2/m_1$ ($m_1$ 
is the mass of the heavier component and
$m_2$ is the lighter secondary component) to calculate $m_2$. 
Further, the first mass could be obtained from the IMF and then calculate the
secondary mass from the $Q$ distribution.  
Since there is a great discrepancy about the actual form of the
$Q$-distribution and how it may depend on the spectral type,
we use the two first ways to create a binary 
although the second way violates the assumptions behind the 
initial IMF derivation. Nevertheless, it is 
interesting to see how the results change if it is used.
If we calculate
the $Q$-distribution by choosing two masses randomly from the IMF, 
for three different
cases IMFs, shown by three histograms together 
with the approximated combined smooth distribution, 
distributions are very similar (figure 2, dot-dashed line).
This can be compared with the observed mass-ratio distribution of
nearby G-dwarf stars 
(histogram and the smoothed distribution, shown in figure 2 as a solid line).
It is understandable that the compatibility is not very good, since 
the observed  $Q$-distribution is for G-type primaries.
As we notice later the PDMF corresponds closely to IMF for low-mass stars and
therefore it has no influence on the discrepancy.
\begin{figure}
\centering
\resizebox{\hsize}{!}{\includegraphics[angle=-90,width=10 cm]{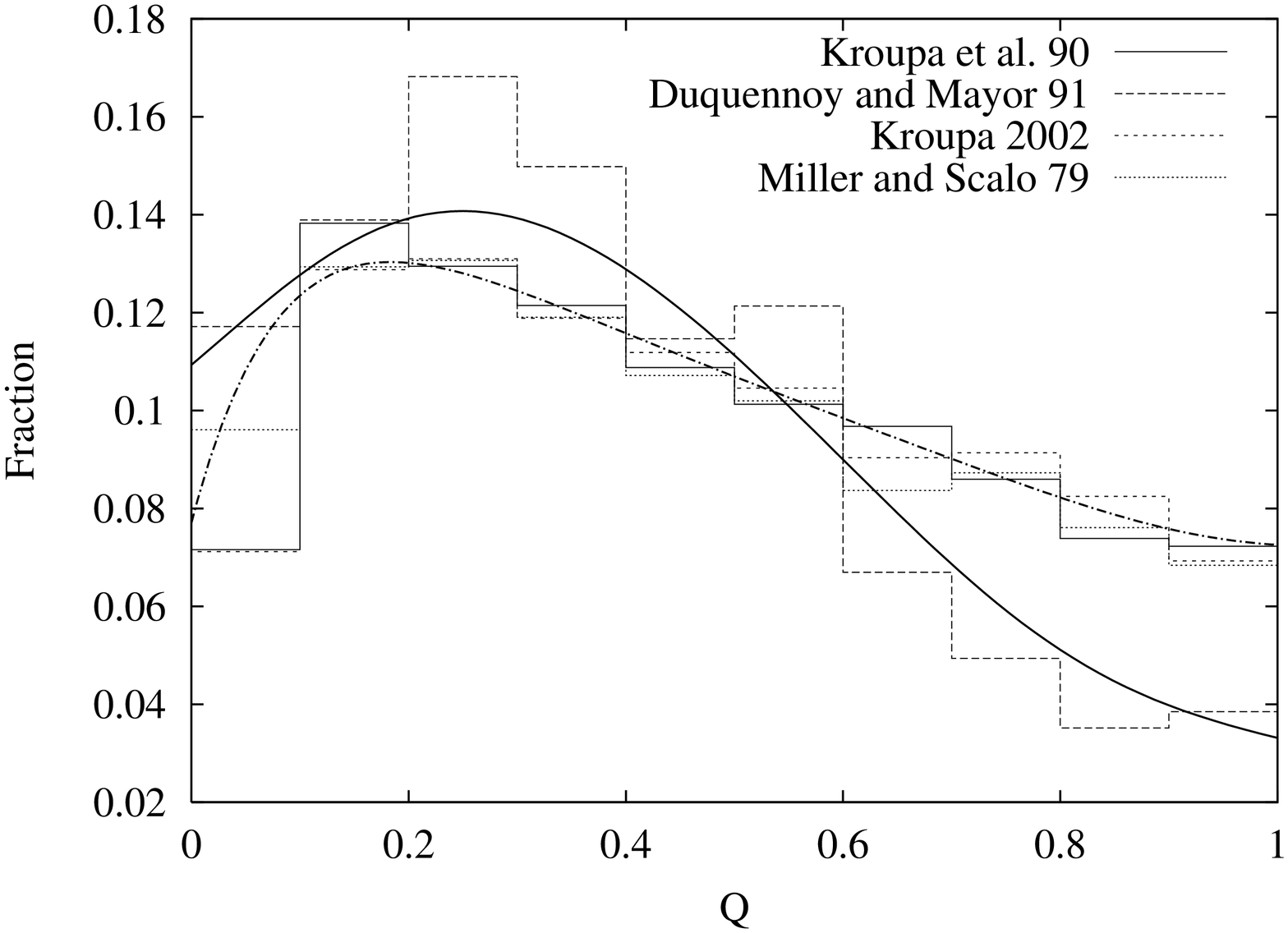}}
\caption{The mass ratio $Q$-distributions as derived from three different
IMF (\cite{miller_79}, \cite{kroupa_90}, \cite{kroupa_02}) and the observed 
$Q$-distribution of G-dwarf stars (\cite{duquennoy_91}).}
\label{}
\end{figure}

\subsubsection{Star formation rate}

Although some studies find evidence of non-uniform star formation rates 
(see, e.g., the review by Rana, 1991), 
for simplicity, we have used a uniform star formation rate
in all the simulations.  
For disk stars we assume that the star formation rate has been uniform between 
10 Myr and 10 Gyr, and for halo stars
between 10 Gyr and 16 Gyr.

\subsubsection{Evolutionary tracks}
Since we start our calculation from the IMF, we have to evolve
stars to derive present-day mass and luminosity functions.
Stellar evolution is done using the analytical stellar evolution 
equations given in Hurley et al. (2000). As an input we
give the present epoch $t_f$ to which the stars are evolved, initial
masses of stars $m_i$ and metallicities $Z$. 
As an output we obtain present-day masses $m_f$, 
effective temperatures
$T_{eff}$ and absolute bolometric magnitudes $M_b$.
In the final output catalogue we have several different types of stars
including main-sequence stars, giants and white dwarfs.
No mutual interaction or mass-transfer between the two stars is included.

\subsection{Galactic distribution of stars}
Since we are interested only in stars having $m_v<15$ we can use a
rather general model for the galactic stellar density.
Galactic disk component clearly dominates the number distributions for 
this magnitude limit even at high galactic latitudes
(see, e.g.,  Ng et al., 1997). This was also confirmed by varying 
galaxy model parameters during the analysis 
and the results remain the same, except for
some extreme unrealistic values. 

Due to  the poor determination of extinction and the resulting large uncertainty of the galaxy
models close to the galactic plane, we have excluded all stars that have
$|b|<10\degr$. 
In our catalogue,
the galaxy model is taken into account by introducing weights to
different stars according to their distances from the earth. 
The weights are related to stellar density functions that are
calculated from Bahcall (1986) where all the functions and model 
parameters are given. Since we use the existing source catalogue we do not
use absolute numbers of stellar counts, 
but instead the weights are used only to correct 
distance distributions of stars. The absolute numbers of stars in a certain
direction $(l,b)$ are taken directly from the GSC catalogue.

\subsubsection{Disk}
The galactic disk density law is a double exponential where the
Galactocentric distance of the Earth is 8.5 kpc and the height above the plane
is 27 pc with local normalization of 0.13 stars per $\mathrm{pc}^3$. 
The scale length is 3.5 kpc and for main-sequence (MS) stars the scale
height depends on the absolute magnitude, $M_{abs}$: $h_z=82.5\times M_{abs}-97.5$ (\cite{bahcall_80}). 
For giants the scale length is constant: $h_z= 250$ pc.
Uniform star formation rate between 10 Myr and 10
Gyr is assumed. Metallicity [Fe/H] is taken from the age-metallicity relationship
given in Rana (1991) and the metallicity distribution is modelled as a
Gaussian with $\sigma=0.2$ and [Fe/H] depends on $t_f$ (\cite{fan_91}). 

	    \subsubsection{Halo}
	    The halo is modelled as a deprojected de Vaucouleurs profile
(\cite{bahcall_86}). The flattening factor is 0.8 and the local normalization is
C=0.00026 stars per $\mathrm{pc}^3$. As for the disk, the star formation rate is assumed
to be constant, but the population II stars are generally much older and 
the age is randomly distributed between 10 Gyr and 16 Gyr. The
age-metallicity relationship for halo stars 
is the same as in Ng et al. (1997) and the metallicities obtained are between
0.0004 and 0.003.

        \subsection{Binary distributions}
The orbital parameters of binaries have been chosen from the following
distributions:
\begin{itemize}
\item{Inclination $i$: uniform in $\cos(i)$}
\item{Eccentricity $e$: in model I, $f(e)=2e$ (i.e. the distribution that is
expected in the case of thermalized orbits due to passing stars
(Hills, 1975) and in model II, we use the multipart
distribution for the eccentricity as given in \cite{duquennoy_91}.
This distribution is derived for G-dwarf stars and may not be valid for
other spectral types, but since they dominate our
sample, we have used it as a second option.
When the orbital period $P<11.6$ d, orbits are circular
and in the intermediate range, $11.6\mathrm{d}<P<1000$d, $f(e)$ is Gaussian
with $\sigma=0.15$ and $\overline{e}=0.3$.
When $P>1000$ d $f(e)=2e$.
}
\item{Semi-major-axis $a$: in model I, $\log(a)$ is assumed to be uniformly
distributed between 0.1 AU and $10^6$ AU (\"Opik 1924). 
In model II, $\log(P)$  has a Gaussian
distribution with $\sigma=0.23$ and $\overline{\log(P)}=4.8$
(\cite{duquennoy_91}).}
\item{Remaining angular elements $\omega$, $\Omega$ and $M$ have a uniform
distribution between 0 and $2\pi$.}
\item{Mass-ratio $Q$: In model a, we choose $m_1$ and $m_2$ independently 
from the same IMF. In model b, the total mass $m_{tot}=m_1+m_2$ is chosen from
the IMF and $m_2$ is calculated using the $Q$-distribution taken from
Duquennoy and Mayor (1991).}
\end{itemize}
Hence, we have four different models $\mathrm{I}_{\mathrm{a}}$, 
$\mathrm{I}_{\mathrm{b}}$ and $\mathrm{II}_{\mathrm{a}}$, 
$\mathrm{II}_{\mathrm{b}}$. In the results we show complete distributions
only for $\mathrm{I}_{\mathrm{a}}$ and for other models we describe how they
differ from the first model.

        \subsection{Simulation algorithm}
Simulation is started by choosing central coordinates for
the field that has a size $A_{field}$. All the objects from the GSC 1.2
catalogue entering the field are
obtained and from these only objects having a star classification are taken.
This underestimates the absolute numbers of stars, since the objects with
non-star classification may be either close-by stars or galaxies.
Since we exclude stars situated at the galactic plane where most non-star
classification objects are
star-star pairs, and at high galactic latitudes non-star objects are most
likely galaxies, we are not introducing any crucial bias to the results.
Then the apparent visual magnitude distribution is calculated for this sample.

After the observational part,
the visual magnitude of the trial star $m_v$ is calculated from 
\begin{equation}
m_v=M_v+5 \log \left( \frac{r}{10 \rm pc} \right ) + A_v,
\end{equation}
where the distance $r$ is taken from a uniform distribution, the absolute
visual magnitude $M_v$ is
obtained from the IMF via stellar evolution algorithms and the extinction
$A_v=3.1 E(B-V)$ is obtained from the colour excess $E(B-V)$ that is
calculated according to the extinction model given by Chen et al. (1999)
that should be a good approximation when $|b|>2.5\degr$.
The colour is obtained using $T_{eff}$ and the data by Flower (1996),
where empirical tables of $T_{eff}$ and  $B-V$ colours are given. 
Also, the bolometric correction $BC$ is calculated
according to Flower's data using the polynomial fits to the tables.
The previous steps are repeated to obtain a visual magnitude 
also to the secondary star.
After visual magnitudes have been obtained for both components 
we generate the 
binary elements and calculate the visual separation $\theta$. 
If $\theta>1.7 \arcsec$ we use the primary magnitude in the fit and
in the case of unresolved binary star we use the combined magnitude.
The procedure is repeated until we have a match between the trial star and a
star in the observed $m_v$ distribution
with the overall accuracy of $\pm 0.5$ mag. 
In this way we fit stellar magnitude
distributions  in our
artificial catalogue to the observed stellar distributions and 
new stars are chosen from the initial sample until all stars in the field
have a match in the distribution.
Then a new field is chosen and the match in magnitude distribution is
obtained in the same way as before.

The
catalogue created in the previous way 
must be corrected since initially the distance is taken 
randomly from the uniform distribution.
The correction is done by weighting every star by the weight 
$\mathrm{w_{i}^{1}}$, that depends on  
the galactic stellar density (calculated using Bahcall (1986) equations)
and $r^2$ since the volume element toward a certain direction
$dV=\omega r^2 dr$, where $\omega$ is the solid cone angle and $r$ is the
distance from the Earth.
We have chosen to use weights just from the practical point of view, since 
the fitting procedure is much faster and 
it is easy to test different galaxy models and the sensitivity of the results to
different parameters by using this approach and not to choose stars from the
real distributions originally.

However, it is not obvious that the magnitude distribution in the catalogue and the
original magnitude distributions are the same after the weight is added to the simulations.
The scaled 
distributions would be exactly the same only if $\mathrm{\sum w_{i}^{1}=f_{m_j \pm dm}}$, 
where $\mathrm{f_{m_j \pm dm}}$ is 
the fraction of stars in the original GSC distribution in the interval $\mathrm{m_j \pm 0.25}$ 
in the jth bin. The index i goes from 1 to the maximum number of stars in the 
corresponding magnitude interval.
Since in every magnitude bin there are samples of stars from different distances, the
previous condition is roughly true. 
In figure 3, we show the original GSC magnitude distribution
together with the distribution obtained from the simulation.
\begin{figure}
\centering
\resizebox{\hsize}{!}{\includegraphics[width=10 cm]{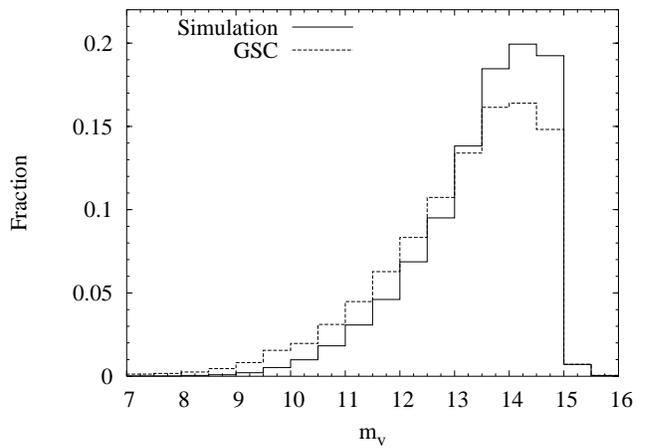}}
\caption{GSC visual magnitude distribution (GSC) as obtained directly from the input catalogue is
shown together with the $m_v$-distribution taken from the simulation.
The simulated distribution deviates systematically from the GSC distribution.
This is corrected by $\mathrm{w_{j}^{2}}$ correction weights.}
\label{}
\end{figure}
We see that there is a systematic change in the new distribution so that 
at small magnitudes, the number of stars is underestimated while at large magnitudes,
it is overestimated. 
To make sure that
this does not cause any distortion to our results, we have corrected for this 
effect. This is done by introducing a new weight $\mathrm{w_{j}^{2}}$ that is the ratio between 
the observed and simulated magnitude distributions. In the final analysis we multiply 
all the stars having a magnitude $\mathrm{m}$ in the interval $\mathrm{m_j \pm 0.25}$ with 
the corresponding additional weight $\mathrm{w_{j}^{2}}$. This guarantees that the final 
magnitude distribution agrees perfectly with the GSC distribution. 
We emphasize that the usage of the additional weight does not change 
the overall distance distribution of stars and 
therefore the influence of the new weight is very small in the final binary distributions.

To make sure that different parts of the simulation have been randomized
enough, we run the simulation a few times using different fields and initial
conditions, but it turns out that the results remain always the same.
Hence, the errors of this study come mainly from the initial conditions and
assumptions used in the simulations and not from the Poisson errors
associated to Monte Carlo simulations.

\section{The analysis of the catalogue}

        \subsection{General properties}
In this section we give statistical distributions of the whole
catalogue without considering any observational effects.
These results represent `true' binary distributions if all of them would
be observed. All real observational projects or astrometric satellites
sample only a small fraction of these populations, depending on the
observational capabilities of the missions. 

For visual binaries there are three main observables $m_{v1}$, $m_{v2}$ and
$\theta$ that observations should be able to provide.  
In figure 4 (left panel) histograms of V-band magnitudes for $m_{v1}$
(brighter component) and $m_{v2}$ are given for Ia and Ib models. 
Also, cumulative distributions are
given (right panel) to make it easy to derive the
fractions of detected stars in different
magnitude limits. The primary distribution cuts sharply at $m_v\approx15$
due to the completeness limit of the GSC and the overall differential magnitude
count distribution $d\Phi(m_v)$ for $m_{v1}$ is presented well by the distribution 
$d\Phi(m_v)\propto 10^{\lambda m_v}dm_v$, where $\lambda=0.25\pm0.03$. 
This distribution is the global average over the whole sky 
and $\lambda$ is different in
different parts of the sky.
Secondary distributions are very wide
and they peak at $m_v=21$ and $m_v=22$ for models b and a, respectively. 
Only 5\% of $m_{v2}$ have $m_v<15$.
In model b, there are more stars having smaller magnitudes, since the mass
ratio distributions are different.
\begin{figure}
\centering
\resizebox{\hsize}{!}{\includegraphics[width=10
cm]{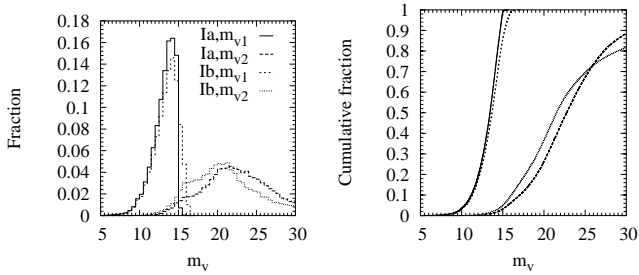}}
\caption{Visual magnitude distribution of primary $m_{v1}$ and secondary
$m_{v2}$ (left panel) and corresponding cumulative distributions (right panel)}
\label{}
\end{figure}

When we calculate magnitude difference  distributions for Ia and
Ib we see that $\Delta m$
has its maximum between 6 and 8 and
that after $\Delta m =3$ the distribution increases quickly (figure 5). 
In model b, the distribution is narrower and drops much more quickly after
the maximum.
The location of 
the  $\Delta m$ maximum value follows mainly from the mass-ratio
distribution and the mass-luminosity relationship.
\begin{figure}
\centering
\resizebox{\hsize}{!}{\includegraphics[width=10 cm]{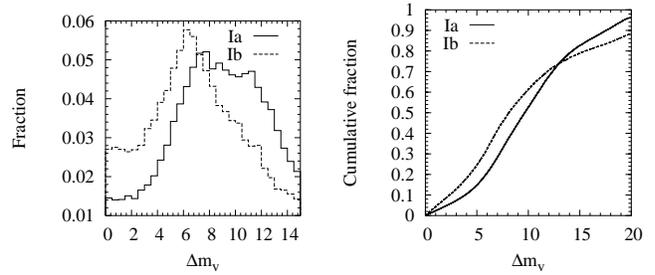}}
\caption{Magnitude difference distributions of binaries for models Ia and Ib 
(left panel) 
and corresponding cumulative distributions (right panel)}
\label{}
\end{figure}

Logarithmic separation, $\log{\theta}$, distributions 
are shown in figure 6,  for models I and II.
Only model a is used, since a and b give similar results.
The distribution of model I is flat up to $\sim 3
\arcmin$, a direct result of the initial $\log{a}$ distribution. Beyond this
the binary frequency drops smoothly and practically no binaries exist with 
$\theta > 50 \arcmin$. 
A gradual decrease in the distribution 
follows from the used upper limit for semi-major axes ($a_{max}=10^6$
AU), 
that causes the upper
limit to $\theta$ and distance beyond which all binaries are unresolved.
Model II distribution is very different from that of model I. 
There are fewer binaries having large separations compared to close-by
systems.
Cumulative distributions for both models are shown in the right
panel and from this we note that half of the objects have $\theta < 1
\arcsec$ in model I and 75\% in model II.   
\begin{figure}
\centering
\resizebox{\hsize}{!}{\includegraphics[width=10 cm]{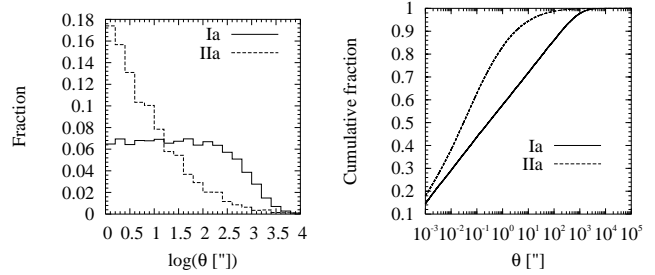}}
\caption{Mutual separation distribution of binary systems (left panel) 
and corresponding cumulative distribution (right panel)}
\label{}
\end{figure}

\subsubsection{Temperature}
Since we have detailed information of the individual stars we can also
calculate the colour B-V histograms and effective temperature histograms for
both components (figures 7 and 9, left panels). Secondaries are typically
much redder than primary stars since fainter stars are typically M-type
stars, but brighter stars are most likely G-type dwarfs. 
In the whole sample 
the most likely
distance is $\sim 700$ pc, but the tail extends to a few kilo-parsecs. 
If only resolvable binaries are included the distance distribution is
naturally very different, since in this distribution the brightness 
is dominated by
primaries that are in unresolved systems.
Color and distance distributions are very similar for all models, the only
difference being that there are more blue secondary stars in model
b than in model a. The difference between models a and b 
is more obvious in effective temperature plot (figure 9, left panel).
In model b, the G-type secondary population is larger.
\begin{figure}
\centering
\resizebox{\hsize}{!}{\includegraphics[width=10
cm]{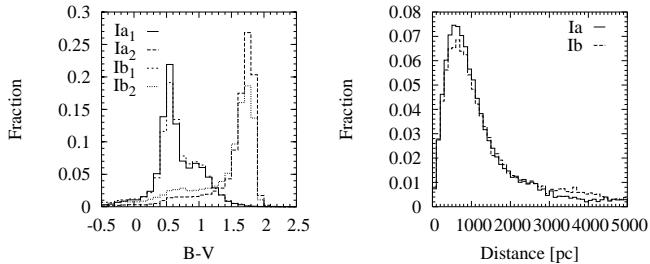}}
\caption{Color B-V distributions for primary and secondary stars (left panel) 
and distance distribution of binaries (right panel). 
Models Ia and Ib are used in the plots.}
\label{}
\end{figure}

\subsubsection{Masses}
In figure 8, initial ($m_i$)  and present-day mass ($m_p$) 
distributions are given 
for both components (model Ia). The primary mass distribution peaks at $1M_{\sun}$ 
and the most common mass for the secondary is $\sim 0.1M_{\sun}$.
Large mass stars with  $M>5 M_{\sun}$ have
evolved to white dwarf secondaries that can be seen in the initial mass
distribution in the right part of the figure. The initial mass distribution
of primaries remains almost unchanged.
Mass distributions of model b differs from the shown histogram, but
looks qualitatively similar and is therefore not shown.

\begin{figure}
\centering
\resizebox{\hsize}{!}{\includegraphics[width=7
cm]{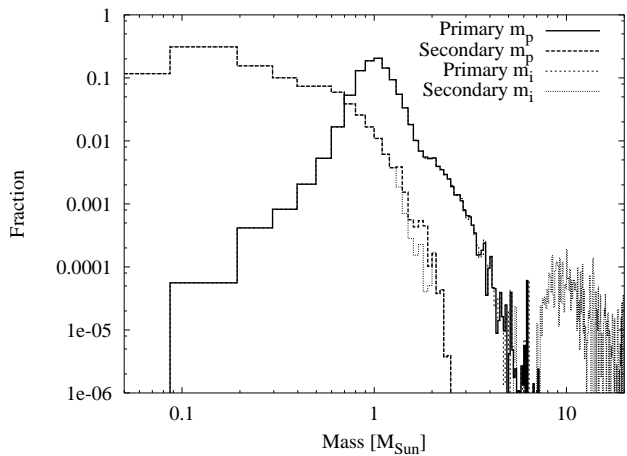}}
\caption{Initial and present-day mass distributions for both components in
model Ia. 
One
solar mass stars dominate the primary distribution and $0.1M_{\sun}$ star
dominates the secondary distribution. Large mass star $M>5 M_{\sun}$ have
evolved to white dwarf secondaries.}
\label{}
\end{figure}

The mass ratio distributions for models Ia and IIb are
plotted in figure 9 (right panel) together with the
calculated distributions using Eq.~(4).
The distirbution of model II differs clearly from the one of model I and 
the calculated distribution does not represent the observed 
G-type binary distribution very well.
If we restrict the sample to G-type stars within 
200 pc from the Earth (figure 10, right panel), then
the fit is better for model b, but for model a the differences are large, when $Q<0.2$. 
The agreement between Duquennoy and Mayor (1991) distribution and model b is
expected, since it was used to generate the initial $Q$-distribution.
In model a, the peak at $0.1<Q<0.2$ comes from the IMF that increases
toward small masses and if we would not restrict our initial masses to
$M>0.08M_{\sun}$, also the last interval $0<Q<0.1$ would be much
higher.
\begin{figure}
\centering
\resizebox{\hsize}{!}{\includegraphics[width=10
cm]{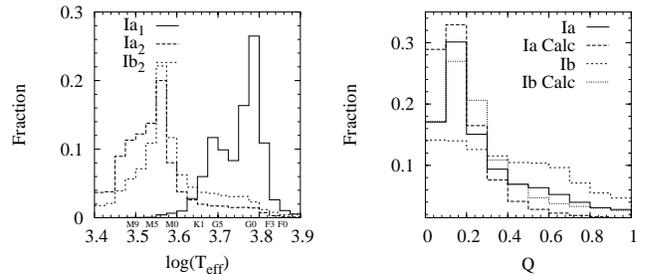}}
\caption{Effective temperature (and spectral type) distributions of all the
stars in the catalogue (left panel). Secondary star  distributions
are shown for models a and b.
In the right panel, mass ratio $Q$
distributions for models Ia and IIb catalogue binaries are given together with
the $Q$ distribution calculated using the equation 4.}
\label{}
\end{figure}

To see how the $Q$-distribution changes as a function of primary mass $M_1$ we
have taken three different mass intervals ($M_1>3M_{\sun}$, 
$1M_{\sun} \geq M_1 \leq 3M_{\sun}$ and 
$M_1<1M_{\sun}$), in which the $Q$-distribution is
calculated separately (figure 10, left panel). 
Since we are interested in qualitative change, we use only model Ia stars.
As expected, the distribution is much more flat for small masses, a fact that
can cause bias effects in some restricted binary samples.
\begin{figure}
\centering
\resizebox{\hsize}{!}{\includegraphics[width=7
cm]{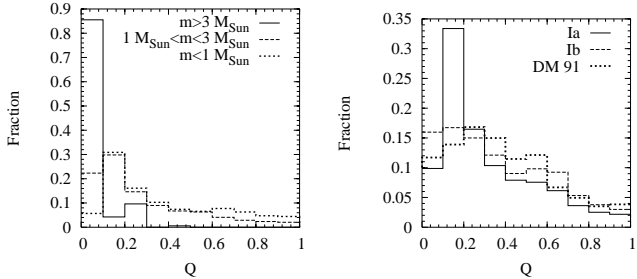}}
\caption{In the left panel, the $Q$-distribution is shown for three different
primary mass intervals (model Ia). In the right panel, we have compared the observed
G-dwarf $Q$-distribution to the ones derived from the model Ia and Ib 
simulations (only
G-type stars are included that have distances $d<200$ pc).}
\label{}
\end{figure}

\subsubsection{Barycentric radial velocities of the components}

Intrinsic tangential and radial velocities of the components
are also of interest for 
astrometric and radial velocity missions. 
To evaluate the intrinsic radial velocity distributions in our catalogue we 
have calculated the barycentric velocity components for both stars.
Tangential velocities are typically very small,
generally less than $10^{-2}$ arcseconds per year and for a typical binary star
the intrinsic tangential motion will not be observed.

Radial velocities can be very large and for distant unresolved binaries it can be 
the only observational method for binary detection. 
Radial velocity distribution is
peaked close to zero and
the distribution is naturally symmetric with respect to the zero axis.
A good knowledge about the distributions can be obtained when
the cumulative distributions are calculated.
The cumulative curves for both binary components are shown in the figure 11.
\begin{figure}
\centering
\resizebox{\hsize}{!}{\includegraphics[width=7
cm]{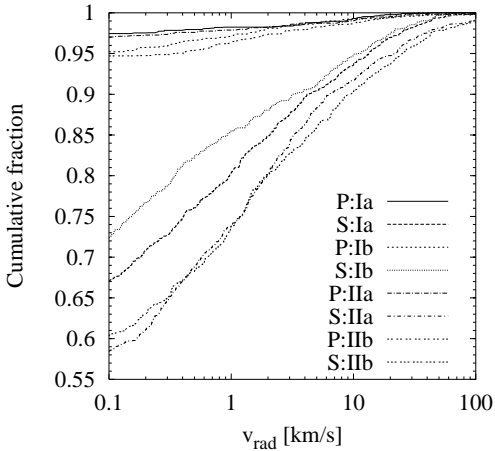}}
\caption{Cumulative radial velocity distributions for both binary components 
given to four different
models.}
\label{}
\end{figure}
We emphasize that the Galactic velocity component is ignored,
since our galaxy model is static.

\subsubsection{Luminosity class}
Main sequence binaries where both components are still in the main-sequence 
is the largest population in all the models. 
Their fraction is 0.68 in model
a and 0.65 in model b.
The second largest group includes
systems  where the primary is evolved away from the MS. 
Their fractions are
0.27 and 0.24 for models a and b, respectively.
Also an interesting sub-group is a pair where the primary is in the
main-sequence and the secondary is a white dwarf.
Their fraction is 0.05 in model a and 0.11 in model b.
The magnitude differences are generally much
larger than for MS+MS binaries. 
The $\Delta m$-distribution starts at $m_v=19$ and peaks at $m_v=26$.
Hence, a typical magnitude difference between a primary and a white dwarf
secondary is $\sim11$ magnitudes.
The typical colour for a white dwarf is
generally bluer than that of the M-type stars. 
These facts makes their detection very difficult, especially for DIVA, where
the CCD is more sensitive to red stars. 
The results derived to white dwarfs 
should be taken only as a first approximation, since no mass
exchange or any other physical interaction between the components is included.
Better modelization of binaries is required if specific close binaries are
to be included in the analysis.

\subsection{Diva observations of the visual binaries}
In principle,  visual double star detection procedures 
can be divided into two different
categories depending on $\Delta m$ and $\theta$. 
If $\Delta m$ is small ($<4$) and $\theta$ is $<1\arcsec$, then the PSF
fitting of the images may show that PSF is a composite of two PSFs.
If $\Delta m$ is large and the separation is well beyond the central PSF 
it may happen that no direct detection of the secondary
is available, but by co-adding all the images at different transits
much fainter background objects can be revealed than would be possible otherwise.
We have developed an efficient procedure that can be used in the image
combination for observations containing binned pixels and rotated images.
The principles of this procedure are given in appendix A.

In DIVA, CCDs
pixels are binned into $1\times4$ pixels in cross-scan direction, the pixel
scale being $0.25\arcsec/$pix $\times$ $1\arcsec/$pix.
Once the images are combined, the initial field of view 
($3\arcsec \times 7 \arcsec$) becomes circular with
a radius of $\sim 3.5 \arcsec$. 
This is the outer limit in our calculations.
The inner limit is set to $1/4$ of the pixel size i.e. 
$\theta_{min}=0.0625\arcsec$.
Beyond the outer limit, only those double stars that are bright enough and 
generate individual windows, can be detected. 
Their number is
statistically very small (see figures 4 and 6) and they are not considered here,
although some of these can be very interesting objects, since accurate
orbits could be calculated. 

We model the observational capability of DIVA when the image
combination procedure is used by writing
that the double star is resolvable if  
	\begin{equation}
	\log{\Delta \theta} > a \Delta m + b,
	\end{equation}
where $\Delta \theta$ is the separation in arcseconds and $\Delta m$ is the difference in
magnitude. The coefficients $a$ and $b$ depend on the procedure by which the
analysis is made. For image combining, the estimated optimal 
values are: $a=0.25$ and
$b=-2$. Also a limiting magnitude must be estimated below which
the signal-to-noise ratio is too small for detection.
Based on our preliminary tests, we define the following
limiting magnitudes for the detection: if $\theta > 1.5 \arcsec$ then
$m_{lim}=22$ and when $\theta \leq 1.5 \arcsec$ then 
$m_{lim}=c\log{\theta}+d$, where $c=-5.5$ and $d=23$.
In reality all the coefficients depend on the number of transits, 
colours of the objects and CCD properties. 
The previous coefficients are for the most common
case, for which the number of images is 142 and the secondary 
star is a red M5-type star.
Also, optimal observing conditions are assumed.

Using the previous conditions we have calculated the fractions of stars that
are observed in the whole synthetic catalogue and show the results in
$\Delta m$ and $\theta$ bins for direct evaluation of the observational
capabilities.
By 'observed' we mean that the stars fullfill the previous 
conditions for $\Delta \theta$ and $\Delta m$ as $m_{lim}$ and 
$\theta$ limits are fixed. 
All fractions are numbers of 'observed' binaries divided 
by the total number of binaries in the whole catalogue.
We show only the histograms for model Ia, since otherwise the plots would be
too confused.

In figures 12 and 13, four curves
are given, where `Combining' means that we have used the previous conditions
directly and in `$m_{lim}=20$' we have set $m_{lim}=20$, but otherwise the
conditions are the same.
The third curve (PSF fitting) is for conservative PSF fitting estimation,
where we still use the previous conditions, but 
we set $m_{lim}=19.5$ and $\Delta m_{max} = 5$.
The last (All) is without
DIVA conditions considering all the binaries with $\Delta m <10$,
$m_{lim}=22$ and $0.0625 \arcsec <\theta< 3.5 \arcsec$.
The plots are different for every model, but model Ia gives the smallest
values and model IIb gives the highest values. The curves for  
models IIa and Ib are located between these two extremes, in this order.
The integrated values of the $\Delta m$ and $\theta$ distributions for all
the models are given in table 1, where we also give values for all stars
that have $0.0625\arcsec <\theta< 3.5 \arcsec$ ($\mathrm{All_2}$) or 
$\theta< 3.5 \arcsec$ ($\mathrm{All_3}$).
\begin{table}
 \centering
 \begin{minipage}{9cm}
  \caption{Fractions of binaries in four different detection categories. 
The last two columns are for the 
whole catalogue when $0.0625\arcsec <\theta< 3.5 \arcsec$ ($\mathrm{All_2}$)
and when $\theta< 3.5 \arcsec$ ($\mathrm{All_3}$).}
  \begin{tabular}{@{}ccccccc@{}}
  Model &Comb&$m_{lim}=20$&$\mathrm{PSF_{fit}}$&All&$\mathrm{All_2}$&$\mathrm{All_3}$\\
   $\mathrm{I_a}$ &0.061& 0.050 & 0.036 & 0.11 & 0.25 & 0.66 \\
   $\mathrm{I_b}$ &0.088& 0.075 & 0.057 & 0.13 & 0.25 & 0.66 \\
   $\mathrm{II_a}$ &0.079 & 0.066 & 0.049 & 0.14 & 0.33 & 0.90 \\
   $\mathrm{II_b}$ &0.11& 0.094 & 0.074 & 0.17 & 0.33 & 0.90 \\
\end{tabular}
\end{minipage}
\end{table}
\begin{figure}
\centering
\resizebox{\hsize}{!}{\includegraphics[width=10
cm]{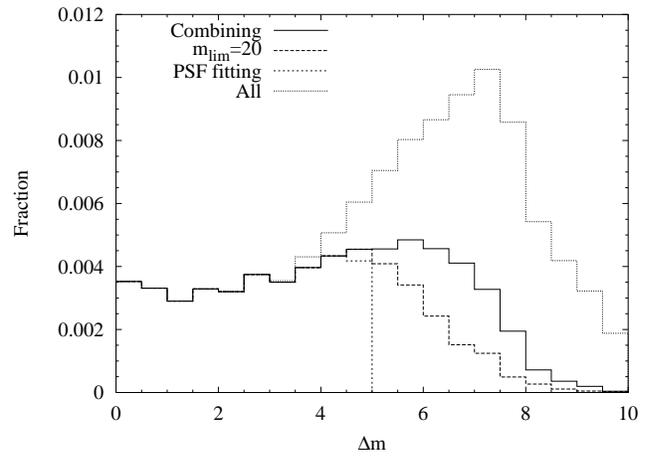}}
\caption{Fractions of binaries having different observability criteria.
Combining means that we use eq. 2 together with  $m_{lim}$ condition 
given in the text. For $m_{lim}=20$ we have used this as a maximum
$m_v$ that could be detected from the images 
and for, PSF fitting, we set $m_{lim}=19.5$
and $\Delta m$ maximum to 5 magnitudes. For 'All' we restrict the sample
to $0.0625\arcsec<\theta<3.5\arcsec$, $m_{v}(\mathrm{secondary})<22$, and 
$\Delta m<10$ without considering any other conditions.
}
\label{}
\end{figure}
\begin{figure}
\centering
\resizebox{\hsize}{!}{\includegraphics[width=10
cm]{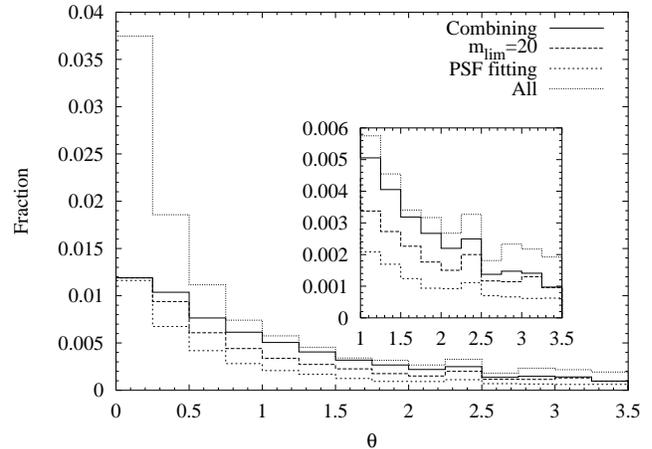}}
\caption{Same as figure 12, but for 8 different separation bins.}
\label{}
\end{figure}
The completeness in each $\Delta m$ or $\theta$ bin for different 
observability criterias
can be calculated if
the corresponding values at the table 1 are divided by the values from the
``All'' column. 
The largest
incompleteness is for small separations $\theta < 0.5 \arcsec$ and for large
magnitude differences $\Delta m >5$. 
Using the image combining method the completeness is between 60\% and 70\%
depending on the model.
The image combination is an
important step to detect large magnitude differences and it can provide a
considerable amount of new double star detections that could not be possible
otherwise.

\section{Properties of `detected' binaries}

        \subsection{Statistical analysis of random pairing}
In the final analysis of binarity in the output catalogue 
it is important to be able to make a distinction between binaries and
optical double stars.
For DIVA this distinction can be
difficult to do if $\Delta m$ is large and no spectral information for the
secondary or the primary from the
dispersed images is available, since for DIVA the sky mapper images are
obtained in white light. 
In these cases we can only
calculate the expected number of background/foreground stars in the field that depend
on the average stellar density down to the magnitude of the secondary star.
Statistically we have $\overline{N}$ background stars in the field, with
$\overline{N}=A_{field}n_{l,b}(m<m_l)$, where $A_{field}$ is the size of the field
and $n_{l,b}(m<m_l)$ is the surface number density of stars having $m<m_l$, $m_l$
denoting the limit down to which stars are calculated and $l,b$ are the
galactic coordinates. 
In a typical case, the field of view is small and 
$\overline{N} << 1$, therefore we have to calculate the probability that we have $N_b$
background stars in the field. If we assume that this follows roughly a
Poisson distribution (see e.g. \cite{scott_89}), then the probability
of having $N_b$ background stars in the field 
is simply 
\begin{equation}
P=\frac{\overline{N}^{N_b}}{N_b !}\exp({-\overline{N}}).
\end{equation}
Respectively, $1-P(N_b=0)$ gives the probability for having at least one
optical double star in the field. 
	
To evaluate the expected number of optical doubles we have calculated the
probabilities using the Bahcall and Soneira (1986) model for stellar density
predictions in three different ecliptic longitudes as a function of
ecliptic latitude (figure 14). To estimate the number of 
stars that are slightly below the maximum detection threshold $m_v(max)=22$
we have used $m_v=23$ in the calculations.
The field used in the calculations is a circle with a radius of $3.5\arcsec$ i.e. 
the one corresponding to the DIVA observational field.
Restricting the sample to only double stars having $\theta<3.5\arcsec$ we exclude 
34\% (in the models Ia and IIa) or 
10\% (in the models Ib and IIb) of all the binaries in our catalogue.
 
The results can be extrapolated to smaller magnitude limits
by dividing the values by
1.4 per magnitude. This estimation is based on the cumulative number distribution 
given by galaxy model predictions.
Hence, only for small galactic latitudes $|b|<20\degr$ and toward the galactic center 
is there a considerable number of
false doubles in the field. 
Beyond this, the probability is very small, of the order of $10^{-2}$.
\begin{figure}
\centering
\resizebox{\hsize}{!}{\includegraphics[angle=-90,width=10
cm]{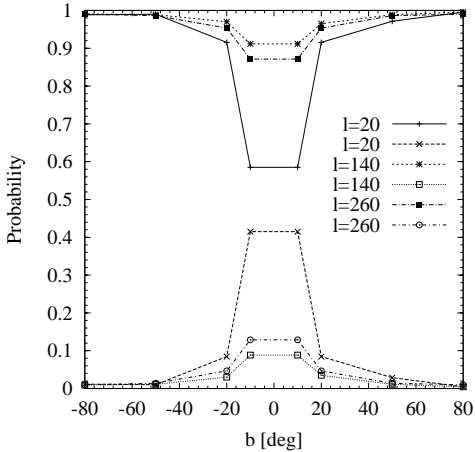}}
\caption{The three upper curves give the probabilities of having no background
stars down to $m_v=23$ in combined DIVA field as a function 
of galactic latitude for three different 
galactic longitudes. The three lower curves give the probabilities of having at
least one optical double star in the field.}
\label{}
\end{figure}	

\subsection{Physical properties}
If we have observations of only one colour or white light observations and
the only observables are $m_1$, $\Delta m$ and $\theta$,
then we can have only a little additional information 
about the  physical properties of the bodies. Some
additional assumptions can allow us to derive more information. If we
assume that the general form of the mass-luminosity relation is 
$L \propto M^{\kappa}$ and suppose that both bodies are on the main sequence, 
then we can calculate the mass ratio directly from the magnitude difference 
\begin{equation}
\log(M_a/M_b)=0.4 \Delta m /\kappa.
\end{equation}
Equation 4 applied to our catalogue gives a rough
estimation for the $Q$ distribution.
We have used $\kappa=4$ in the calculation  
(figure 9, right panel). 
However, there is clearly a trend that 
the equation underestimates mass ratios at large values and
overestimate $Q$ at small values.

In addition, if we know spectral types of the stars and parallax estimations 
we can calculate  absolute magnitudes to the components. 
This can give us
the upper limit to the age of the system if the primary is a massive star 
and hence exclude the ambiguity
for the secondary to be either a low mass MS star or a white dwarf, 
even if the
spectral type of the secondary is not known.
For bright binaries, parallax and proper motion 
measurements can resolve the background star
ambiguity if the measurements are accurate enough. 
If a binary is a nearby system we might see a trail in the combined
image revealing the motion of the secondary. 
However, the fraction of these objects is
very small,  but they might be interesting cases for later
binary studies, since the orbital elements could probably be determined.  

\section{Discussion}
Our synthetic catalogue and the 
analysis of double star observations include 
several approximations.
Especially important to the output is the way binaries are assumed to form. 
The results in this study agree with the results of Duquennoy and Mayor
(1991). In a recent study, Halbwachs et al. (2003) found that
the $Q$-distribution is very different from the Duquennoy and Mayor
(1991) distribution, but only for binaries with $P<50$ days. This does not
change our results, since only a very small fraction of binaries in our
sample have $P<50$ days (7\% in models I and 9\% in models II). 
Also, we believe that the uncertainties associated
with the galaxy model or binary distributions are not that severe and the
results represent statistically correct binary distributions.
A separate but closely related topic is to estimate the number of astrometric
binaries and how these overlap with visual binaries.
It is important to know how different types of binary detection methods are
complementary  and how the global binary analysis should be done. This
is closely related to the details of  astrometric missions and is out
of the scope of this study.

In the future we will do similar studies, but applicable to GAIA, using a
more accurate galaxy model and including binary evolution in the 
simulations. This step is crucial to GAIA, since the magnitude limit is much larger. 
However, some
results may not change remarkably for visual binaries since the
resolvability criteria restricts the sample within a few kpc.
Since multiband photometry is done by GAIA, much better physical information
can be obtained for both components and better estimations for binary
properties of the system can be made. 
	
\begin{acknowledgements}
This project was supported by the ESA-PRODEX contract
"From HIPPARCOS to GAIA" with no. 14847/00/NL/SFe(IC).
Dr. J. Hurley is gratefully acknowledged for providing the routines of his
analytic stellar evolution code and the anonymous referee is acknowledged for
valuable comments that led to the improvement of the paper.
\end{acknowledgements}

\appendix
\section{Double star detections with large magnitude differences by co-adding
images at different transits}

If an image combination procedure must be applied to every
star, that has typically $\sim 10^2$ transits during the astrometric mission
with probably several different images from each transit, 
then the procedure must be simple and fast, while still being able
to enhance the resolution and to increase the signal-to-noise ratio in the combined
image.
These requirements in mind we have developed an image combining procedure 
that is based on the drizzling algorithm developed
by Fruchter and Hook (2002). 
The drizzling or the 
method of variable-pixel linear reconstruction is a procedure
in which input pixels (for DIVA 1x4) are sub-sampled to some output grid that
have smaller 1x1 pixels (by a fraction $f_{1}$) than in the initial frame.
Before actual mapping, the initial pixels are
shrunk (by a fraction $f_{2}$) or some other procedure (e.g. Gaussian weighting)
is used so that the final image is not smoothed too much (Gilliland et al.
1999). 
Different weights
$w_{xy}$ can be introduced to individual pixels so that 
geometrical distortions or any relevant corrections can be taken into account
if we have a priori knowledge about them. 
The kernel of the
algorithm, that is repeated until all the images are combined, is
\begin{eqnarray}
W_{xy}^{`}=a_{xy}w_{xy}+W_{xy}\\
I_{xy}^{`}=\frac{a_{xy}i_{xy}w_{xy}s^2+I_{xy}W_{xy}}{W_{xy}^{`}},
\end{eqnarray}
where $I_{xy}$ is the combined intensity, $W_{xy}$ is the combined weight, $w_{xy}$ is
individual weight of the pixel, $a_{xy}$ is the geometrical weight (overlapping
factor), $s$ is the fraction between the size of the output pixel to
the input pixel and, finally,
$i_{xy}$ is the measured pixel value. 
In equations (A.1) and (A.2), $x$ and $y$ refer to pixel positions in the
input and output grids and dotted values refer to values at the next iteration.
The free parameters $f_{1} \leq1$
and $f_{2} \leq1$ must be determined from general algorithm tests.
The modification of the basic pixel drizzling procedure 
is related only to the way the geometrical overlappings are
calculated. For binned pixels we can either subdivide the pixels back to
$1\times 1$ pixels that are individually shrank or then another possibility
is to shrink $1\times 4$ pixels directly by some fraction before they are
mapped to the new grid.

\subsection{Different phases in the image combination and object detection}

The main steps in the procedure are:

\begin{enumerate}
\item  Shift and rotate

\item  Choose new background grid by ratio $s=pix_{out}/pix_{in}$ finer than
the grid in the original CCD

\item  Shrink initial pixels to reduce grid convolution effects (by factor $f_{2}$)

\item  Map shrank pixels to underlying grid and add all images together using
the modified drizzle algorithm

\item  Create the PSF for the central hypothetical single star using all the
information that is available and affects the PSF 
(i.e. positions, magnitude, colour etc.) 

\item  Run PSF through the same algorithm using the same $f_{1}$ and $f_{2}$
values 
as used in the main frame containing
additional components

\item  Scale images and subtract the combined PSF from the main frame
calculated in step 4.

\item  Apply an identification algorithm to the final image to detect faint companions
\end{enumerate}

The last four steps are required to detect anything with large $\Delta
m$ close to the central component and they all are normally performed, 
since we generally do not know what is
hidden in the images. 

In an ideal case after the adding procedure 
the signal-to-noise ratio for $i$ images is
\begin{equation}
\frac{S}{N}=\frac{\sum S_{i} }{\sqrt{\sum N_{i}^{2}}}%
\end{equation}

The ideal magnitude gain can be easily estimated.
If all the images are similar $S_{i}=S_{j}=S$ and only Poisson error is
considered, then in $n$ images we get a magnitude threshold higher by
\begin{equation}
\Delta m_{gain}=2.5\log\sqrt n.
\end{equation}
This makes a difference of 1 mag in 6 images, 1.25 for 10, 1.85 for 30 and
2.22 for 60 images. However, in reality, this is not exactly true and
especially  for faint objects the signal is readout noise limited and then
\begin{equation}
\frac{S}{N}\approx \sqrt{n} 
\sqrt{\left( \frac{S}{1+R^2/S} \right)},
\end{equation}
where $R$ is the readout noise. Hence, the proportional factor is $<2.5$, but
the gain is still proportional to $\sqrt n$.

In reality the noise behavior is more complicated and the drizzling
procedure itself has an effect on the noise since, in adjacent pixels,
the noise is correlated. Due to the correlation, the measured noise in a
drizzled image is smaller than that obtained by just interlacing the images. A
detailed noise behavior depends on $f_{1}$, $f_{2}$ and overlapping geometry.

\subsection{About PSF subtraction}

The main purpose of PSF subtraction is to see faint features around bright
objects. To do good PSF fitting and subtraction, high S/N is required and 
good knowledge
about the center and the wings of the PSF is required. For a small number
of rotations, binning can be a severe problem in the
detection capability, 
since the magnitude gain can be small 
and unequally distributed
outside the central area. 
If the PSF changes as a function of position on the chip and the
spectrum of the target, or if there is any change in the PSF shape in time 
(focus etc.) that is not known a priori, then
subtraction with the ``wrong'' PSF 
can create some spurious errors and false detections. One
possibility to solve this problem would be to use PSFs obtained at 
close-by locations and to build up a PSF library from where PSFs can be
obtained. 
Also, simulated PSFs could be used but this does not solve the problem
of a PSF evolving with time.

\end{document}